# The labour and resource use requirements of a good life for all

Chris McElroy[1,2], Daniel W. O'Neill[2,3]

[1] Community Development and Applied Economics, University of Vermont, Burlington, Vermont, United States of America
[2] Sustainability Research Institute, School of Earth and Environment, University of Leeds, Leeds, LS2 9JT, United Kingdom
[3] UB School of Economics, Universitat de Barcelona, C/ de John Maynard Keynes 1-11, 08034 Barcelona, Spain

## Abstract

We use multi-regional input–output analysis to calculate the paid labour, energy, emissions, and material use required to provide basic needs for all people. We calculate two different low-consumption scenarios, using the UK as a case study: (1) a "decent living" scenario, which includes only the bare necessities, and (2) a "good life" scenario, based on the minimum living standards demanded by UK residents. We compare the resulting footprints to the current footprint of the UK, and to the footprints of the US, China, India, and a global average. Labour footprints are disaggregated by sector, skill level, and region of origin.

We find that neither low-consumption scenario provides a realistic path to providing a good life for all. While the decent living scenario would require only an 18-hour working week, and on a per capita basis, 35 GJ of energy use, 4.0 tonnes of emissions, and 5.5 tonnes of materials per year, it fails to provide essential needs. The good life scenario encompasses these needs, but would require a 46-hour working week, 73 GJ of energy use, 7.5 tonnes of emissions, and 13.2 tonnes of materials per capita. Both scenarios represent substantial reductions from the UK's current labour footprint of 65 hours per week, which the UK is only able to sustain by importing a substantial portion of its labour from other countries. We conclude that limiting consumption to the level of basic needs is not enough to achieve sustainability. Substantial changes to provisioning systems are also required.

# 1. Introduction

## 1.1 Providing for human needs within planetary boundaries

Current global provisioning systems are both unsustainable and ineffective, leaving billions without basic necessities while contributing to a worsening environmental crisis (Kikstra et al., 2021). Moreover, no country currently meets the basic needs of its residents at a level of resource use that is globally sustainable (O'Neill et al., 2018). Researchers have repeatedly shown that necessary reductions in resource use are not compatible with an ever-increasing gross domestic product (GDP) (Haberl et al., 2020; Parrique et al., 2019). Yet reductions in resource use must be made compatible with meeting basic needs.

Gough (2015) proposes using basic and universal requirements for human beings to provide a foundation for achieving justice within environmental limits. To that end, Doyal and Gough (1991) list universal characteristics of need satisfiers that they suggest are common to all cultures, including adequate food and water, health care, physical security, and basic education. Rao and Min (2018) define a set of decent living standards based on this list and several other wellbeing theories. These decent living standards provide a basis for calculating the minimum resource use and emissions associated with meeting basic needs. Millward-Hopkins et al. (2020), for example, use Rao and Min's (2018) standard and find that decent living standards could be met globally with 60–75% less energy consumption than at present. Druckman and Jackson (2010) apply a UK-specific minimum income standard established by Bradshaw et al. (2008), and show that this standard could be met with 37% lower emissions than the UK had at the time of their study.

Such analyses offer a way to understand both the potential and limits of low-consumption approaches to sustainability. However, there is as yet no similar evaluation of the minimum labour requirements in the literature. This article aims to fill this gap in knowledge, both building on — and further developing — the living standards used by Millward-Hopkins et al. (2020) and Druckman and Jackson (2010) to evaluate the paid labour requirements of achieving a good life for all people.

## 1.2 Research aims and approach

This article seeks to answer three primary research questions:

1. What are the paid labour requirements of providing a good life for all?
2. What are the energy, emissions, and material requirements of providing a good life for all?
3. How do these requirements compare to the current footprints of different countries and to sustainability thresholds?

A good life for all is modelled through two scenarios: a "decent living" scenario based strictly on the decent living standards proposed by Rao and Min (2018), and a "good life" scenario based on a more generous minimum income standard published by the Joseph Rowntree Foundation (Hirsch, 2019), which is based on the deliberations of UK focus groups. These scenarios model reduced consumption compared to the UK baseline, and assume that the existing makeup of the global economy remains the same. The analysis centres on the UK, as it is the basis for the minimum income standard used by Druckman and Jackson (2010), and it provides a good case study of the scope to reduce labour footprints. The labour requirements are specifically compared to the existing footprints of the UK, US, China, India, and the global average. Energy, emissions, and material footprints are included to both contextualise the results and demonstrate the environmental impact of each scenario. While the calculations are based on UK consumption and its existing supply chains, the results are scaled to represent general per capita footprints for each scenario.

The remainder of this article proceeds as follows. Section 2 provides more background on the existing minimum requirements literature. Section 3 provides a detailed account of each scenario and the methods used to calculate the footprints. Section 4 presents the results for both low-consumption scenarios and all footprint variables. Section 5 discusses the implications of the results, their limitations, the potential for future research, and the novel contributions of this article. Section 6 concludes.

# 2. Background

## 2.1 Minimum resource requirements

Several researchers have estimated the minimum resources required to meet basic needs, including the required energy, emissions, and materials. Estimates of the minimum energy required to meet basic needs extend as far back as 1985, when Goldemberg et al. (1985) argued that the basic needs of far more people could be met without increasing energy consumption in developing countries. Goldemberg et al. (1985) concluded that 33 gigajoules per capita per year (GJ/cap/year) would be sufficient. Rao and Baer (2012) developed an early version of the decent living standards framework for minimum emissions and energy calculations, which Lamb and Rao (2015) improved upon and used. Lamb and Rao (2015) predict that countries in Africa and Asia could meet basic needs with between 28–40 GJ/cap/year. More recent bottom–up estimates based on Rao and Min's (2018) decent living standards have found much lower values. Rao et al. (2019) estimate that 11–24 GJ/cap/year would be sufficient for decent living in Brazil, India, and South Africa. This result is also in line with a study by Millward-Hopkins et al. (2020), who estimate a requirement of 15 GJ/cap/year globally, ranging from 13–18 GJ/cap/year for individual countries. Kikstra et al. (2021) report an average requirement of 17 GJ/cap/year globally, with an even larger range of 9–36 GJ/cap/year in different regions.

While some estimates of the energy required to meet basic needs are as high as 100 GJ/cap/year (Rao et al., 2019), these tend to either assume that social and economic systems will change relatively little (e.g. Arto et al., 2016), or they provide higher standards of living for countries that are already better off (e.g. Grubler et al., 2018; IEA, 2019, p. 90). Millward-Hopkins et al. (2020), by contrast, only include regional variations based on differences in demographics and geography, not existing differences in wealth or living standards, and calculate the energy required to meet specific needs instead of extrapolating from correlations between quality of life and energy demand. Kikstra et al. (2021) also apply the same living standards universally, but account for existing differences in urbanisation, infrastructure, and diets, thus leading to the larger range in their results.

Studies based on Rao and Min's (2018) decent living standards yield far lower estimates of the minimum energy necessary to meet basic needs in part because the basic needs they consider are quite restricted. Rao and Min's (2018) standards feature minimal housing space, only 9 years of education, one phone and one television per household, and vehicles and washing machines that are shared between households. The other standard used in our analysis, the minimum income standard for the UK published by the Joseph Rowntree Foundation (Hirsch, 2019), is much more generous, including larger apartments for couples and families, basic furniture and cooking utensils, and small amounts of income for occasionally dining out or traveling within the UK.

In addition to energy, emissions and material requirements have been studied as well. Druckman and Jackson (2010) find that the 2008 version of the same minimum income standards (i.e. Bradshaw et al., 2008) would lead to per capita yearly emissions of 17 tonnes of $CO_2$ equivalent. Lamb and Rao (2015) find that only 4.1–6.7 tonnes of $CO_2$ equivalent per person are likely required to meet basic needs and provide a high life expectancy. Huo et al. (2023) report that emerging economies could achieve decent living standards with only 14.7 Gt of yearly emissions, which corresponds to 2.3 tonnes per person per year. Akenji et al. (2019) calculate the carbon footprints associated with different countries' lifestyles, and find that they range from 10.4 tonnes $CO_2$ equivalent per person per year in Finland, to 2.0 tonnes in India. Vélez-Henao and Pauliuk (2023) find that the material footprint of the decent living standards proposed by Rao and Min (2018) would be between 2 and 11 tonnes per person per year (depending on the exact implementation), with 5 tonnes for their reference scenario.

## 2.2 Minimum labour requirements

Several researchers have suggested that reductions in working time have the potential to improve wellbeing and reduce carbon emissions (Coote et al., 2010; Kallis et al., 2013; Stronge and Harper, 2019). However, despite its potential value, there are few examples in the literature concerning the minimum labour requirements (i.e. the maximum possible reductions in working time) that mirror the energy research discussed in Section 2.1. Many authors have predicted the potential for working-time reductions over the past 150 years, often arguing that working weeks could be reduced to between 15 and 25 hours (e.g. Lafargue, 1883; Keynes, 1930; Gorz, 1994; Bregman, 2017). However, this argument is largely based on extrapolations of historical improvements in average labour productivity, alongside the potential benefits of shorter working weeks, rather than a consumption-based analysis of the labour that is required to meet people's needs. These predictions have also largely failed to materialise due to the economic incentive to maintain employment, as explained by Jackson (2017): in our current economic system, productivity gains tend to lead to increased production — a larger GDP — rather than greater leisure. This tendency represents a version of the Jevons paradox based on labour efficiency (Polimeni et al., 2008).

Many countries in the global North have already found a way to reduce their labour requirements significantly — by offshoring the production of goods and services. For example, according to Coote et al. (2010), working-aged people in the UK average 19.6 hours of paid work per week (including students and others who are not working)[1]. These same people enjoy the products of more than double their own labour, given that the majority of UK labour is imported. Sakai et al. (2017) calculate the UK's labour footprint based on the products and services purchased by UK residents. They find that only 40% of the labour embodied in the UK's consumption is domestic (Sakai et al., 2017). The UK's increased leisure, then, reflects its ability to extract labour from other countries more than it represents a realisation of Keynes' vision of a 15-hour working week for his generation's grandchildren (Keynes, 1930).

At present, few examples exist of consumption-based analyses of minimum labour requirements. Hardt et al. (2020) provide a description of the structural changes required for a "post-growth" economy, but leave the determination of the necessary labour levels in such an economy to future research. Kallis (2017) points out that at least some industrial labour seems necessary, but he does not provide a more detailed analysis. Popper-Lynkeus argued in 1912 that basic needs could be met with a 12-year, 35-hour-per-week labour service (Martinez Alier, 1992). However, the economy has changed dramatically over the past 113 years, and understandings of the material requirements of a good life have changed with it (Davis et al., 2022; Grubler et al., 2018). In summary, there is a clear need for a consumption-based, sector-specific analysis of the minimum labour required to provide a good life for all people in a modern economy.

[1] This is about 27.4 hours per week equivalent, using the units we use for our own results, which account for time off work (see Section 3.2).

# 3. Methods

## 3.1 Scenarios

Within our analysis, we explore two scenarios that aim to meet basic needs with minimal overall consumption: (1) a "decent living" scenario based on the decent living standards of Rao and Min (2018), and (2) a "good life" scenario based on the minimum income standard published by the Joseph Rowntree Foundation (Hirsch, 2019). Throughout our analysis we refer to these as "low-consumption" scenarios, as they represent a reduction in consumption compared to the UK baseline, and the goal of the scenarios is to meet basic needs at a low (i.e. sustainable) level of consumption. However, these scenarios represent an increase in consumption for many countries in the world, where basic needs are not currently being met.

The per-category spending in each scenario, along with the recorded UK spending in 2019 (i.e. the baseline spending), is shown in Table 1. This section provides an overview of the two scenarios, while more detailed descriptions are provided in Supplementary Materials 1.

**Table 1. Consumption in the UK in each scenario, by spending category.** Values are in millions of 2019 euros (to match EXIOBASE 3 data).

| Categories | Baseline spending | Decent living spending | Good life spending |
|---|---:|---:|---:|
| Groceries (food and drinks) | € 96,128 | € 116,791 | € 132,025 |
| Clothing | € 39,800 | € 38,979 | € 38,979 |
| Housing | € 341,038 | € 153,052 | € 190,547 |
| Utilities and insurance | € 99,848 | € 41,950 | € 44,852 |
| Healthcare | € 274,245 | € 35,270 | € 191,971 |
| Appliances, furnishing, and maintenance | € 106,977 | € 0 | € 32,949 |
| Education | € 139,680 | € 76,260 | € 139,680 |
| Devices (TVs, phones, computers) | € 47,809 | € 11,384 | € 27,159 |
| Transport | € 254,257 | € 101,951 | € 101,951 |
| Public administration and defence | € 206,364 | € 47,884 | € 149,197 |
| Recreation (vacations, toys, dining out) | € 477,114 | € 0 | € 124,416 |
| Care work (childcare, senior care) | € 43,853 | € 0 | € 36,147 |
| Gross fixed capital formation | € 434,246 | € 0 | € 385,353 |
| *Total* | *€ 2,561,359* | *€ 623,522* | *€ 1,595,227* |

### 3.1.1 The decent living scenario

The decent living scenario was constructed to include, as closely as possible, the household and collective requirements specified by Rao and Min (2018) in their "decent living standards". This scenario therefore permits direct comparison with other studies that draw on these standards, such as the energy studies of Rao et al. (2019) and Millward-Hopkins et al. (2020). Rao and Min (2018) include 12 household requirements and 10 collective requirements in their decent living standards. These are included within the 13 spending categories used in our analysis (Table 1). Clothing and transport budgets were kept at the same levels as the minimum income standards defined by Hirsch (2019). Spending in the remaining 11 categories was reduced below these levels to reflect Rao and Min's (2018) requirements.

Many of these reductions are based on explicit exclusions by Rao and Min (2018). For example, although the scenario includes groceries, clothing, heating and cooling equipment, basic household appliances, and a mobile phone for each adult, it does not include alcohol, dining out, or personal computers. Rao and Min (2018) limit education to 9 years, and annual healthcare spending to PPP $700 per person. These values leave education spending at 55% of the baseline scenario and healthcare spending at 13% of the baseline scenario. The decent living standard includes a minimum floor space of 30 $m^2$ for each family, with 10 $m^2$ of additional space for each person after the third (Rao and Min, 2018). Given the size of UK apartments (Roberts-Hughes, 2011), this means that essentially all families would be living in single-bedroom-sized apartments, and this is reflected in the housing budget. Diets are based on those budgeted by Hirsch (2019). While they are not plant-based, which tend to have lower footprints (Gephart et al., 2016), they were reviewed for nutritional adequacy, and this is in line with the requirements specified by Rao and Min (2018). Many other categories are not discussed by Rao and Min (2018), such as furniture (e.g. mattresses, eating/cooking utensils, chairs), legislative and judicial bodies, police or fire departments, care work, maintenance of built capital, and insurance. As such, spending on these is also set to zero. For a full description of the decent living scenario, please see Supplementary Materials 1.

### 3.1.2 The good life scenario

Within the UK, the Joseph Rowntree Foundation has used group discussions to determine a set of minimum living requirements for UK residents, and converted these into a minimum income standard. This research has been carried out for 17 years (e.g. by Bradshaw et al., 2008; Davis et al. 2012; 2018; 2022; Hirsch, 2019), to continually establish what UK residents consider "the basic or default minimum standard, below which it is unacceptable for anyone to have to live" (Bradshaw et al., 2008, p. 16). This minimum income standard, as published by Hirsch (2019), forms the basis of our good life scenario.

The good life scenario has similar goals to the decent living scenario, but aims to include many important things that are excluded from the decent living scenario, such as the government, university education, maintenance of built capital, recreation, and greater healthcare spending. The budgets allocated by Hirsch (2019) were left unchanged for every category except for care work, housing, and healthcare, and three categories that we added (education, public administration and defence, and gross fixed capital formation). The care work budget was set to 20% of the Hirsch (2019) spending numbers because Hirsch (2019) assumes that all of a child's parents work full time, but in practice this is only true for about 20% of children (Bradshaw et al., 2008). The housing budget was extended to include the cost of providing social housing. The healthcare budget provided by Hirsch (2019) also does not include public spending on healthcare, so we instead used the UK's overall baseline healthcare spending, with a 30% reduction based on the repeated finding that about 20–40% of healthcare spending is wasted (WHO, 2014; Berwick and Hackbarth, 2012).

Importantly, the Hirsch (2019) budgets do not include education, public administration and defence, or gross fixed capital formation (GFCF). Education spending was kept at the baseline level, as a considerable amount of both new research and skill development seems necessary to support this scenario. All baseline public administration costs were included except for those relating to defence. GFCF spending was scaled to include only what is required to cover the costs of depreciation (i.e. to maintain the stock of built capital). For a full description of the good life scenario, please see Supplementary Materials 1.

## 3.2 Multi-regional input–output analysis

We used multi-regional input–output (MRIO) analysis to determine the paid labour footprints associated with the consumption profiles for the two low-consumption scenarios. MRIO analysis requires input–output tables that link the sales of each sector, in each region, to every other sector. We used EXIOBASE 3 (Rasul et al., 2024; Stadler et al., 2025), a set of input–output tables designed for environmentally-extended MRIO analysis[2]. This method involves inverting the tables to reveal the resulting production in all sectors, and from all countries, from a single unit of consumption in any particular sector (Kitzes, 2013). The resulting production values are then scaled by the impact of one unit of production in each sector to determine the total impact of any unit of consumption (Kitzes, 2013). This approach allows for the evaluation of any consumption scenario across a wide range of impacts, including the use of particular resources, the emission of different substances, or the labour required for production (Stadler et al., 2018). However, MRIO also has its limitations. Since it is a macroeconomic technique based at the sectoral level, it can only be used to calculate resource and labour footprints at this level. Although we considered using other techniques such as life-cycle analysis, which allow footprints to be calculated at the product level, we chose to use MRIO as it is a particularly good method for accounting for international trade flows, which are important within our analysis.

To present more understandable and extensible results, we present labour footprints as the hours of labour per week equivalent, by dividing the total hours of required labour by the number of working weeks per year, the working age (15–64 year-old) population of the country, and the percent of working years we expect one person to work. UK law requires 5.6 weeks per year for holiday (GOV.UK, 2023), so our calculations assume about 46.6 working weeks per year worked. We assume that 80% of a person's potential working years are spent working (40 out of 50 years), which is both in line with global employment figures and allows time for essential activities such as attending school and raising children. The working age population is based on data from the OECD (2024), although for consistency the total population numbers for the low-consumption scenarios are based on the UK household data that were used to generate the scenarios. Since the conversions are all linear, alternate results can be derived through multiplication with preferred factors.

[2] Our labour account uses data that will be released with EXIOBASE version 3.9.6. All other data are based on EXIOBASE version 3.9.5.

Footprints for our two low-consumption scenarios were compared against footprints based on the real consumption of the UK, US, China, India, and the global average. The year 2019 was chosen as the year for our analysis to use the most recent non-extrapolated data available from EXIOBASE 3, while avoiding the effects of the global pandemic in 2020. We disaggregated the resulting labour footprint in each scenario by consumption category, sector, origin, and skill-level. Sectors were aggregated into seven categories based on the concordance tables provided by Wood (2022). Energy, emissions, and material footprints were calculated for each scenario and each country. The emissions indicator that we use is the AR5 GWP100 indicator, which measures greenhouse gas emissions in $CO_2$-equivalent terms based on their global warming potential. The energy indicator that we use is final energy, which represents the energy delivered to consumers, ready for use. The material footprint indicator that we use is raw material consumption, which accounts for the total mass of biomass, fossil fuels, metal ores, and non-metal ores extracted globally and used within a country. Since changes in direct use are beyond the scope of MRIO analysis, the footprints assume that energy and emissions associated with direct use scale proportionally with changes to embedded energy and emissions footprints. Each of these scenarios, including the baselines used for comparison, focuses on total consumption, so changes in inventories and changes in valuables, which do not involve consumption, are omitted. We used the *Pymrio* library (Stadler, 2021) to conduct our analysis.

## 3.3 Scenario alignment

The underlying data for the two low-consumption scenarios feature categories that are distinct from each other, and from the product sectors in EXIOBASE 3. To allow for alignment between these sources, final demand was separated into a custom set of 13 categories (Table 1). Data from Hirsch (2019) were central to this alignment process, and therefore our 13 categories most closely match the 17 categories used by Hirsch (2019). Of these, six categories were merged with others, one was excluded (Council tax), and three were added (education, public administration and defence, and gross fixed capital formation). Categories were merged for overall simplicity and for more tractable calculations. For example, the "social participation" category used by Hirsch (2019) includes funds for a TV, laptop, gifts, stationary, domestic travel, a passport, and donations (Loughborough

University, 2023b). These funds were split into two separate categories (i.e. recreation and devices), based on detailed accounts from Loughborough University (2023b). Council tax was excluded as it does not count as final demand. The three new categories were added to better account for non-household spending.

Calculating alternate spending totals required a comprehensive account of household types and costs. Household costs in the UK in 2019 were based on the estimates from Hirsch (2019), using detailed breakdowns provided by Loughborough University (2023a). These per-household costs were then multiplied by the number of households of each type, based on values from the UK Office of National Statistics (ONS, 2023). Because these accounts have different sets of household types, the Hirsch (2019) household budgets were adjusted to fit the types used by the ONS (2023) before multiplying them to produce a full weekly UK budget for each category. These adjustments were based on demographic data from the ONS (2020; 2023).

To align changes in these budgets with the MRIO accounts, the 200 sectors used by EXIOBASE 3 were sorted into our 13 categories (shown in Table 1). This concordance process was largely based on the sector definitions provided in the concordance tables from Wood (2022), with further clarification based on classification definitions from the UN (2018) and category definitions from Bradshaw et al. (2008), Davis et al. (2012; 2018; 2022), and Hirsch (2019). The full concordance table is available in Supplementary Materials 2. Note that many sectors with zero final demand in the UK in 2019 were not included in the concordance. Changes in demand for each category were modelled as a proportional increase or decrease in spending for all sectors within that category. Good fits for spending types were prioritised over matching total spending per category, to preserve the makeup of each category when scaled. Because of this approach, the grocery category has higher spending in the low-consumption scenarios than the baseline (see Table 1). This result reflects a difference in where spending is allocated, and not a true increase in consumption.

Since non-profit and government spending are concentrated into just a few sectors in EXIOBASE, both types of spending were consolidated with household spending into a single spending vector. Gross fixed capital formation (GFCF) was kept as a separate vector, denoted by its total under the category with the same name. Changes to total GFCF were used to scale GFCF spending for each sector proportionally.

To scale government spending, public spending data were sourced from HM Treasury (2019). The EXIOBASE total for UK spending on public administration and defence in 2019 is within 5% of public spending reported by HM Treasury (2019), when excluding components that are accounted for elsewhere or that are not final spending (i.e. education, healthcare, waste management, recreation, public debt transactions, social protection, and EU transactions). This pruned-down version of public spending was then used as a basis for scaling the EXIOBASE spending account, proportional to the resulting change any cuts would have in reported spending. Definitions for each category were drawn from Eurostat (2019). The full calculations are included in Supplementary Materials 2.

# 4. Results

## 4.1 Overall labour requirements

The UK labour footprint in 2019 was 64.7 hours per week equivalent, while the decent living scenario footprint is 17.5 hours per week equivalent, and the good life scenario footprint is 45.5 hours per week equivalent (Fig. 1). These footprints represent the total amount of global labour embodied in the direct and indirect production of the goods and services purchased in the final spending amounts provided in Table 1. The totals do not include any labour or spending on exports, except as necessary for final domestic consumption. As discussed in Section 3.2, these equivalent units distribute the work over a population the size of the working age (i.e. 15–64 year-old) population of the UK in 2019, and assume that people work 80% of their working age years, with 5.6 weeks per year spent on holiday.

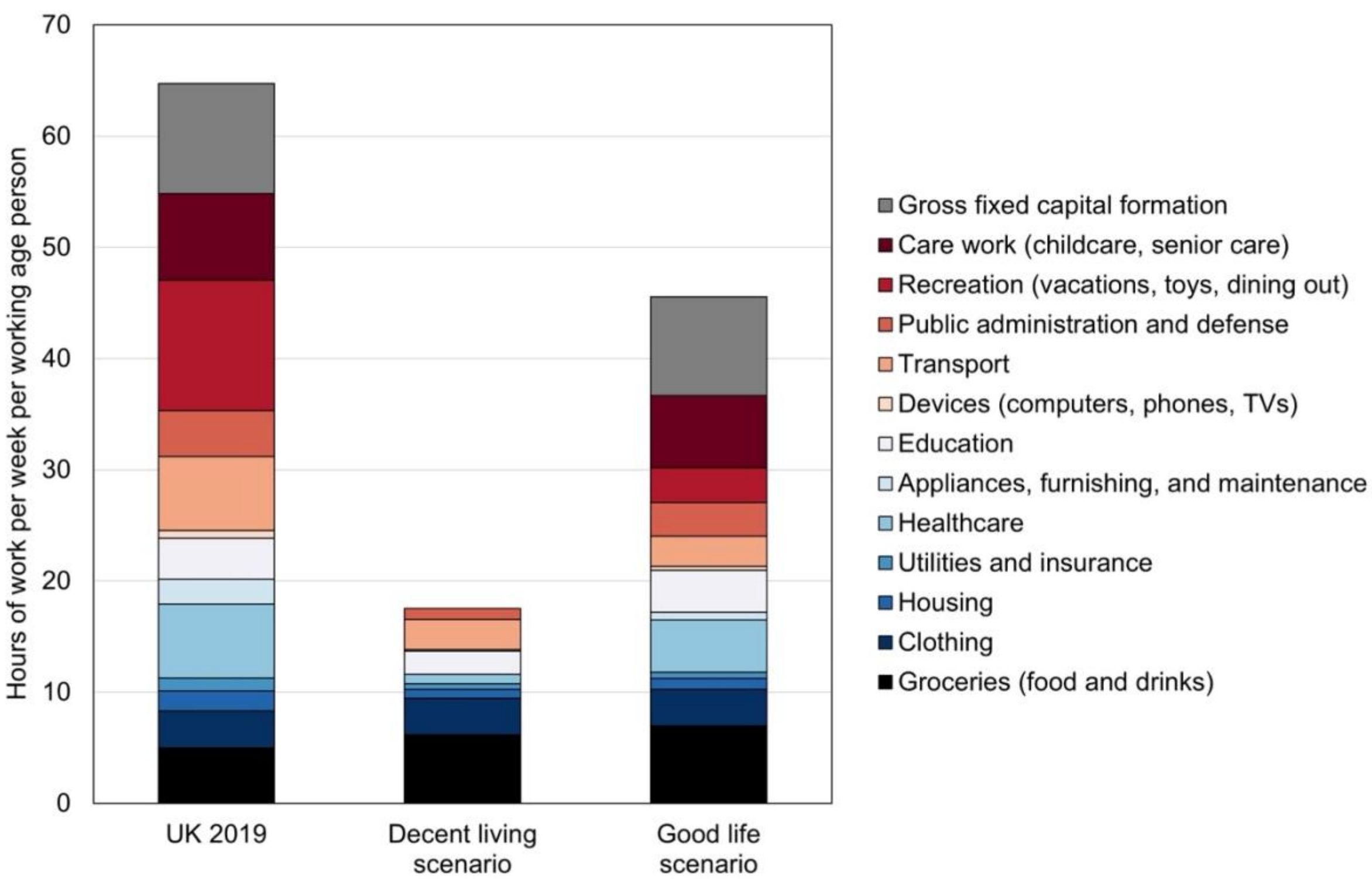

**Fig 1. Labour footprint for each scenario, split by spending category.** Results are reported in hours of work per working week, divided by 80% of the working age population (ages 15–64) of each region in 2019. Results are split by the categories of spending that drive each hour of labour (not types of labour).

The good life scenario represents a reduction in labour footprint of 19.2 hours per week compared to the baseline, while the decent living scenario represents a further reduction of 28.0 hours per week. About 66% of this latter reduction is due to the three categories that are entirely absent from the decent living scenario: recreation, care work, and maintenance of the stock of built capital (GFCF). The rest is primarily the result of increased spending on public goods, with public administration, healthcare, and education alone accounting for 80% of the remaining difference. Labour for groceries is higher in both the low-consumption scenarios versus the baseline scenario, as a result of the higher spending in this sector discussed in Section 3.3. These apparent increases are offset by decreases in food and drinks purchased at stores and restaurants, which mostly fall under recreation in baseline spending. The main categories separating the good life scenario from the baseline scenario are recreation (8.6 hours per week less in the good life scenario) and transport (3.9 hours per week less). The 11 other categories combined represent a difference of 6.6 hours per week.

## 4.2 Labour compositions and comparative analysis

The labour footprints of the UK and US are far higher than the labour footprints of China, India, and the global average (Fig. 2). The US has a labour footprint of 66.6 hours per week equivalent, 1.9 hours per week above the UK, while China, India, and the global average have footprints that are all more than 20 hours per week below the UK's. India's labour footprint is 31.7 hours per week, China's is 43.5 hours per week, and the global average is 37.7 hours per week.

The country comparisons also help contextualise the labour footprint for the two low-consumption scenarios. The labour footprint for the decent living scenario is substantially below the global average, and 1.8 times smaller than India's. Conversely, the good life scenario has a labour footprint that is 7.9 hours per week above the global average. Thus while it is a low-consumption scenario with respect to the UK and US, it is not a low-consumption scenario globally.

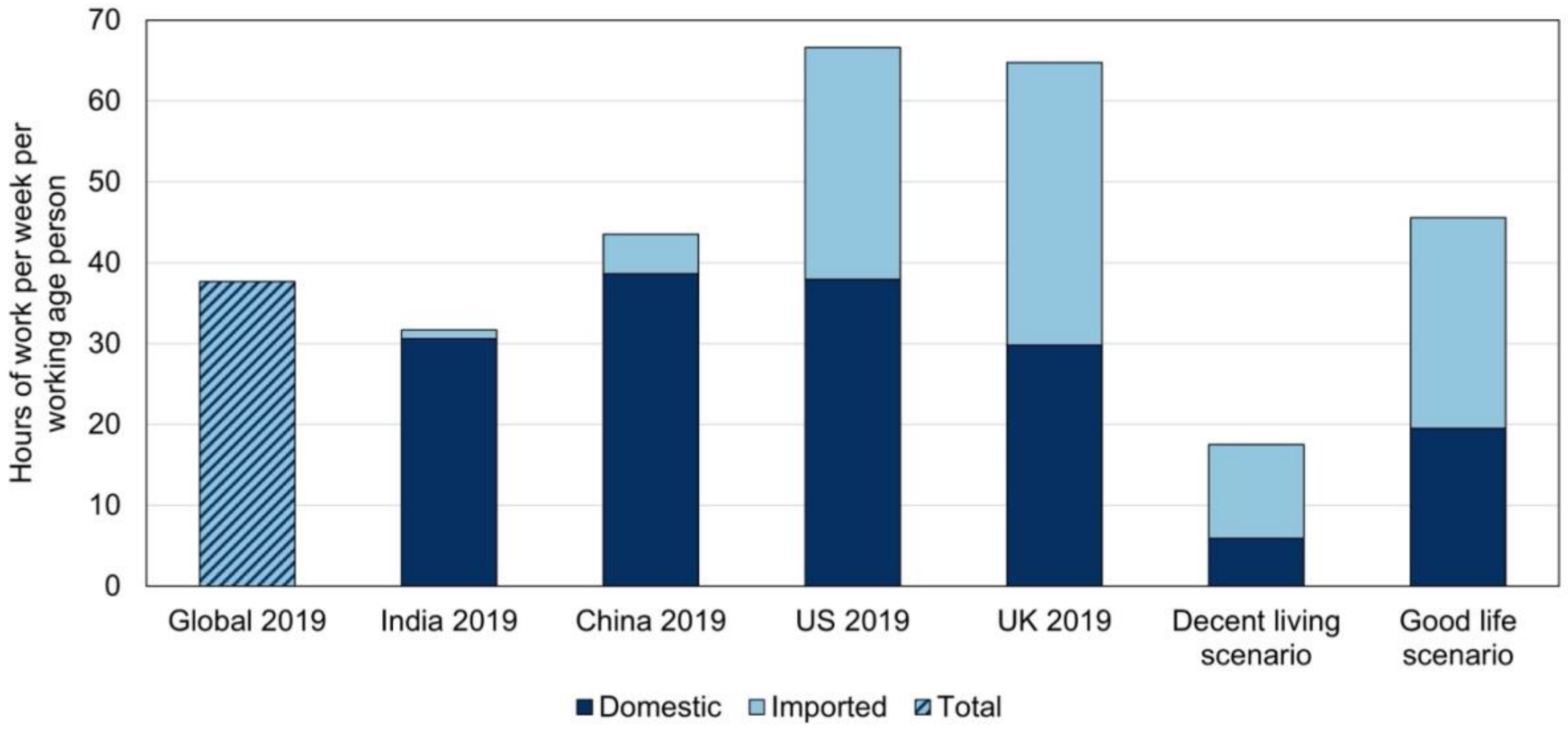

**Fig 2. Labour footprint for different countries in comparison to the two low-consumption scenarios, split by source of embodied labour.** Results are reported in hours of work per working week, divided by 80% of the working age population (ages 15–64) of each region in 2019. Totals are split into domestic and imported labour.

A key part of the story of global labour is the labour embodied in imports (Fig. 2). There is relatively little difference between each country's domestic labour footprint (that is, the labour in each country that goes towards producing products that are consumed in that country). The UK has the lowest domestic labour footprint (30 hours per week) while China has the largest (39 hours per week). Most of the difference between labour footprints arises from the different amounts of embodied labour each country imports from other countries. UK residents, for example, import labour equivalent to 110% of the total (domestic and imported) per capita labour footprint of Indian residents. This imported labour accounts for 54% of the UK's labour footprint, while India imports only 3% of the labour in its footprint.

The fraction of labour imported in the two low-consumption scenarios is quite similar to the fraction observed in the baseline UK footprint, although the decent living scenario includes a higher proportion of imports. The decent living scenario has a higher proportion of imports because it features less spending in sectors with a high proportion of domestic labour, such as healthcare and care work. More generally, though, the relatively similar fraction of domestic versus imported labour across the UK scenarios is a product of the method used in our analysis, which assumes that the UK's relative spending between regions for a given consumption category remains unchanged. For example, no allowance was made for the possible prioritisation of domestic production over foreign production, to avoid making the unrealistic assumption that sectors in different countries produce identical goods. As a result, however, the composition of each low-consumption scenario is primarily a reflection of existing UK supply chains.

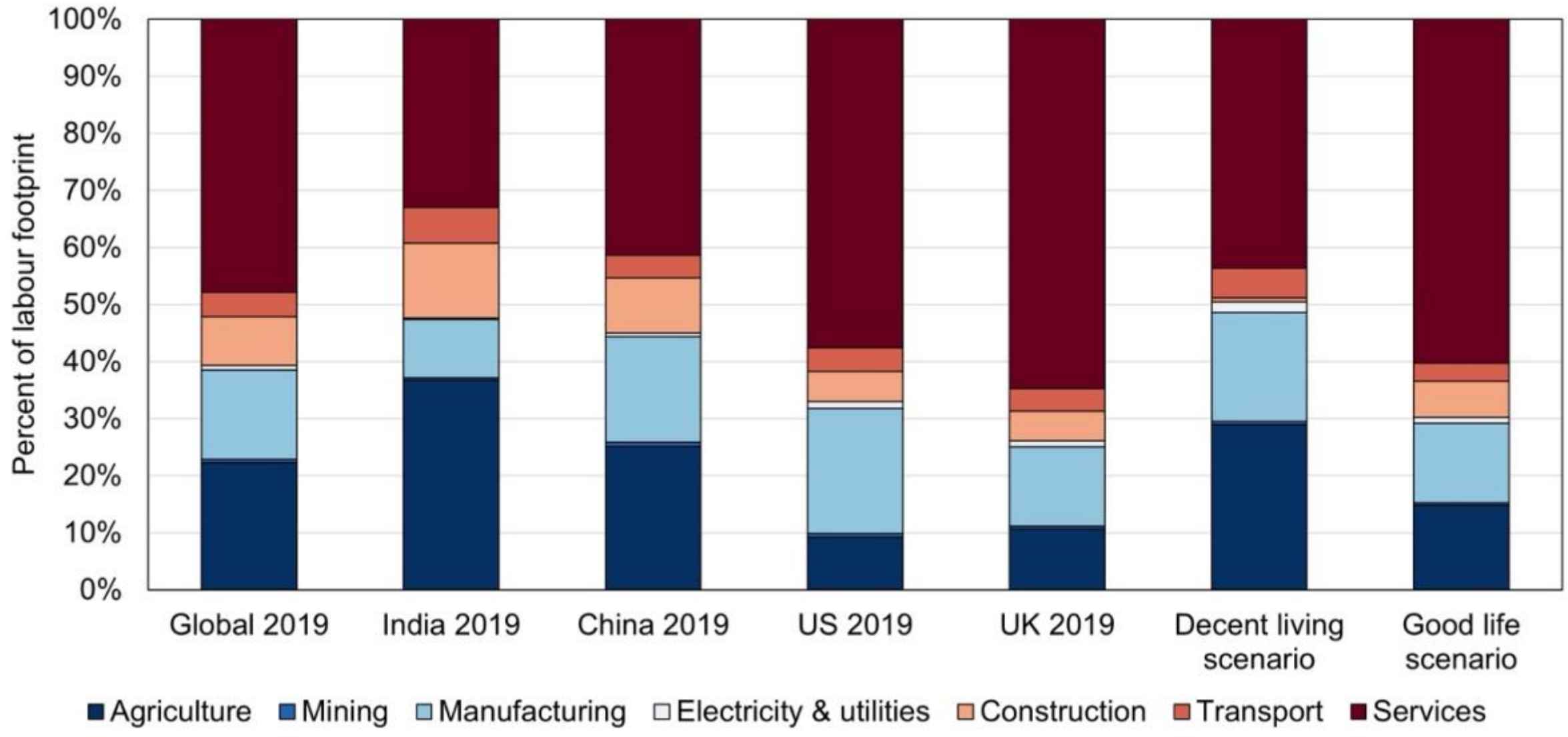

**Fig 3. Sectoral composition of each labour footprint.** Compositions are reported as a percentage of the overall footprints shown in Fig. 2. Note that changes in inventory are not included as part of demand. Proportions are by hours of work, not by number of workers.

By contrast, the sectoral composition differs markedly between the two low-consumption scenarios, and the international comparisons provide quite helpful context (Fig. 3). Interestingly, the sectoral makeup of the labour footprint for the decent living scenario is very similar to that of China; both include 25–30% agriculture, 15–20% manufacturing, and 40–45% services (although China includes much more labour in construction). By contrast, the good life scenario's sectoral composition is quite similar to the UK baseline, with 10–15% of labour in agriculture and manufacturing, and 60–65% in services.

Both of the low-consumption scenarios feature more agricultural work, in proportional terms, than the UK baseline. This difference is most pronounced for the decent living scenario, in which agriculture makes up 29% of total labour. For comparison, the labour footprints of both the UK and the US feature less than 11% agriculture (and over 55% services). The large differences in relative agricultural work may be misleading, however, since the absolute differences are quite small. The UK, US, and both low-consumption scenarios include between 5 and 7 hours per week equivalent of agricultural work. China and India's footprints include about twice the amount of agricultural labour, at 11 and 12 hours respectively.

The differences in skill level composition are less noticeable, but the international comparisons again serve as helpful context (Fig. 4). The proportion of different skill levels required in both low-consumption scenarios is similar to the UK baseline. The proportion of low-skilled work is also similar to the global average (13%), but the two low-consumption scenarios require a higher proportion of high-skilled work (around 33%), and a lower proportion of medium-skilled work (around 54%), than the global average. The proportion of high-skilled work in the low-consumption scenarios is roughly twice that of India and China. The relatively high percentage of high-skilled labour in the decent living scenario is arguably not consistent with the fact that it only includes 9 years of education, so high-skilled labour would be difficult to provide in this scenario. That said, given the low total number of hours of work in the decent living scenario, it only requires about half the labour in each skill type than is available globally.

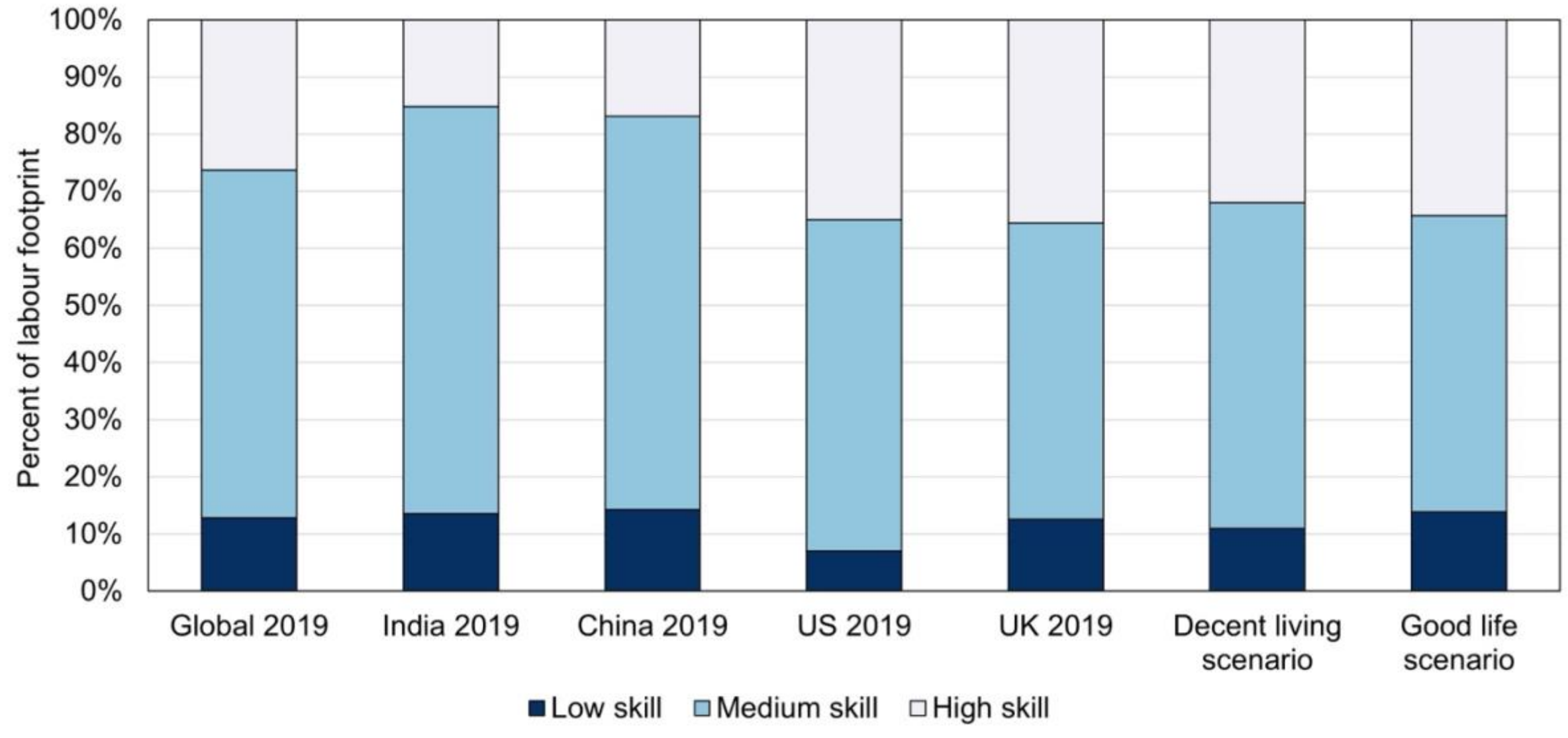

**Fig 4. Skill level composition of each labour footprint.** Compositions are reported as a percentage of the overall footprints shown in Fig. 2. Note that changes in inventory are not included as part of demand. Proportions are by hours of work, not by number of workers.

## 4.3 Energy, emissions, and material requirements

The low-consumption scenarios reduce the energy, emissions, and material footprints of the UK (Fig. 5). The final energy footprint of the UK in 2019 was 117 GJ per person per year. The decent living scenario uses only 30% of this footprint, while the good life scenario requires 62% (Fig. 5a). The ratio of imported to domestic energy use in the good life scenario is 1.7:1, similar to the baseline UK ratio of 1.6:1. The decent living scenario includes much more imported energy, in proportional terms, with a ratio of 2.3:1. The final energy required for the decent living scenario (35 GJ/cap/year) is far below that of the UK, US, China, and the global average, although still well above India's 18 GJ/cap/year footprint. The final energy footprint of the good life scenario (73 GJ/cap/year) exceeds all but the footprints of the UK and US. Notably, while the US has a similar labour footprint to the UK, its energy footprint is 114% higher. This difference is due to the much higher domestic energy footprint in the US (a finding which also holds for emissions and materials). The direct use of energy varies roughly proportionally with each country's overall energy use, validating the proportional estimate used in our analysis.

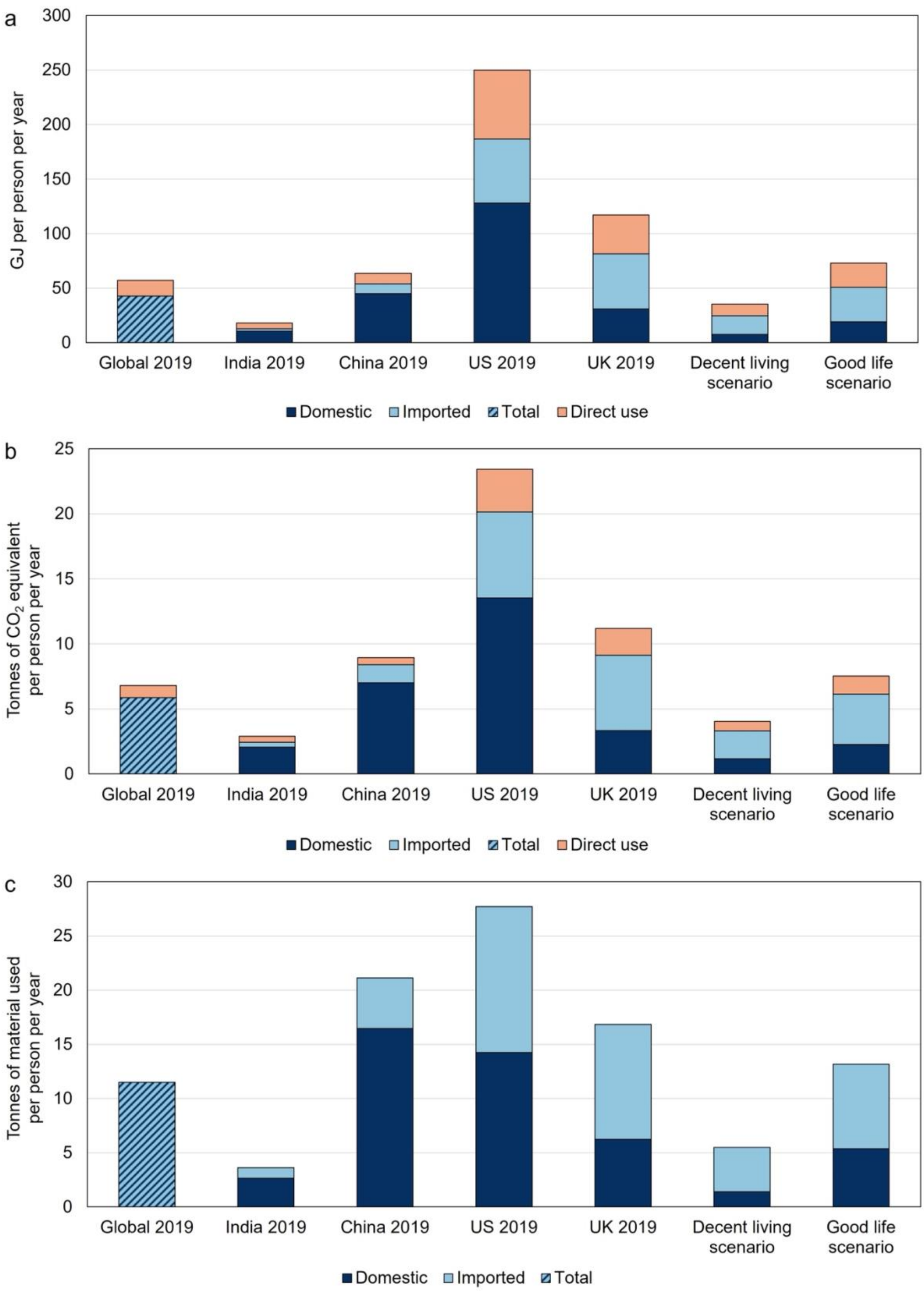

**Fig 5. Resource footprints for (a) energy, (b) emissions, and (c) materials, split by source.** All values are per capita per year, based on 2019 population data. Results are split into domestic, imported, and direct use. Direct use in the low-consumption scenarios is scaled to match reductions in embodied amounts.

The emissions footprints (Fig. 5b) are quite similar to the energy footprints, although the emissions in the low-consumption scenarios are slightly higher relative to the baseline scenario. The decent living scenario has 4.0 tonnes of $CO_2$-equivalent emissions per person per year, compared to the baseline of 11.2 tonnes per person per year, while the good life scenario has emissions of 7.5 tonnes per person per year. The US, again, far exceeds all other countries' footprints at 23.4 tonnes, while India is responsible for 12% of the US's per capita emissions, at 2.9 tonnes. The decent living scenario emissions are 41% lower than the global average, while the good life scenario represents a 10% increase. As with energy, the direct emissions caused by residential heating, transportation, and other non-industrial sources scale roughly proportionally with each country's overall emissions.

The material footprint results (Fig. 5c), which show raw material consumption, paint a similar picture. The baseline requires 16.8 tonnes per capita, the good life scenario 13.2 tonnes per capita, and the decent living scenario 5.5 tonnes per capita. The ratios of imported to domestic material consumption are very similar to the ratios for energy, at 1.7:1, 1.5:1, and 3.0:1 for the baseline, good life, and decent living scenarios respectively. In contrast, China, India, and the US all have ratios under 1:1. Clearly, UK spending on sectors with high levels of foreign material consumption is mirrored in both low-consumption scenarios.

# 5. Discussion

## 5.1 Scenario feasibility

Overall, our results suggest that it is not feasible to achieve a good life for all people with a socially sustainable labour footprint, within the scenarios that we have considered. Providing a good life for all would require workers to work 45.5 hours a week, approaching the UK's 48-hour limit on weekly working time (The Working Time Regulations, 1998). Generalising this lifestyle to other countries would require increasing global working hours by 7.9 per week. The decent living scenario would involve a much more sustainable 17.5 hours per week of work, but the lower hours in this scenario are only made possible by excluding many important societal needs.

For example, the decent living scenario does not include funds for the executive or legislative bodies of government, public safety, universities, childcare, insurance, or maintenance of built capital. We argue that for a scenario to be realistic, it should include the government, safety, skilled workers, children, risk, and built capital. Together these categories make up 68% of the gap between the good life scenario and the decent living scenario; including them would increase the working hours in the decent living scenario to 36.7 hours per week.

Some of the consumption reductions within the decent living scenario seem harsh to the point of infeasibility. Healthcare spending in the UK would be reduced to only 13% of its baseline amount. The UK healthcare system is already struggling with a lack of funding (Montgomery et al., 2017); this change would serve as a drastic impediment to care. At home, people would have to survive without beds, furniture, eating utensils, or other basic goods. Each household, no matter its number of inhabitants, would live in a one-bedroom-sized apartment. There would be no alcohol consumption, and no toys, personal computers, or other leisure goods. While individuals might be living, their life does not seem very decent. Rao and Min (2018), who proposed the decent living standards framework, themselves note that these standards are "instrumental (but not sufficient) to achieve… wellbeing" (p. 231). This is why we proposed the good life scenario as a more sufficient alternative.

Beyond the findings on social sustainability, our results also suggest that the environmental footprints of the two scenarios are not sustainable. For a 67% chance of limiting global warming to 2 °C, the global carbon budget between 2020 and 2100 is 1150 Gt $CO_2$, or about 1.75 tonnes per person per year if distributed evenly over a population of 8.2 billion people (IPCC, 2023). The values associated with the decent living and good life scenarios are much higher, at 4.0 and 7.5 tonnes per person, respectively. For reference, existing commitments would leave the global average at about 4.3 tonnes per person per year by 2030 (Alcaraz et al., 2018). Although a sustainability threshold for material footprint is harder to quantify than $CO_2$, previous authors have estimated a value of around 7 tonnes per person per year (Bringezu et al., 2015; O'Neill et al., 2018; Fanning et al., 2022). While the material footprint of the decent living scenario is below this threshold, at 5.5 tonnes per person, the good life scenario clearly exceeds it, at 13.2 tonnes per person per year.

## 5.2 Implicit imperialism

In some ways, the good life scenario demonstrates the extent to which imperialist appropriation is baked into the social norms and laws of the UK. The budgets provided by Hirsch (2019) represent the minimum acceptable standard for residents of the UK. Yet the footprints implied by this standard require labour, emissions, energy, and material footprints much higher than the global average. Given that humanity is already transgressing six planetary boundaries (Richardson et al., 2023), a scenario that involves increasing emissions, material extraction, and working hours is clearly not sustainable. The UK enjoys a high standard of living because it is able to appropriate many goods and services from the rest of the world, while allowing its residents to live a life of relative leisure.

If a good life is to be achieved for all people, then the living standards demanded by a given country should match the labour supplied by that country. At present there is a large mismatch, with countries in the global North demanding a higher standard of living than they are willing to sustain through their own labour. This imbalance adds another dimension to the drain from the global South that previous authors have identified (e.g. Hickel et al., 2022). While global trade is not inherently unsustainable, a society that relies on asymmetrical appropriation of goods from other countries cannot serve as a model for sustainability.

## 5.3 Comparison with previous literature

Several authors have predicted that working weeks could be reduced to 15–25 hours in the future (e.g. Lafargue, 1883; Keynes, 1930; Gorz, 1994; Bregman, 2017). The predictions of these previous studies are all substantially lower than the 45.5-hour working week of our good life scenario, although they are in line with the 17.5-hour working week of the decent living scenario. The difference between these predictions and our good life scenario largely stems from the changing social standards of the UK over time. The resulting increases in living standards are now seen as essential, as acknowledged by both the Joseph Rowntree Foundation (Davis et al., 2018) and Rao and Min (2018).

As an illustration, we can look to Popper-Lynkeus's 1912 prediction of a 12-year, 35-hour-per-week labour service (Martinez Alier, 1992). To make this prediction directly comparable to our results, we need to stretch the labour service over 40 years, and include 5.6 weeks of vacation per year. The result is a 12-hour working week, well below our good life scenario. However, Popper-Lynkeus's prediction does not include the labour needed for modern healthcare, computers, transport, or appliances — all of which are seen as necessities today.

Druckman and Jackson (2010) report that their minimum income scenario would lead to 37% lower $CO_2$ emissions for the UK (compared to their baseline year of 2004). The good life scenario, based on similar household budgets but extended to include other important societal needs, yields 33% lower emissions than the baseline value (7.5 tonnes versus 11.2 tonnes per capita). The decent living scenario yields a far larger (64%) reduction in emissions (to 4.0 tonnes per capita), although it represents much steeper cuts in living standards than those modelled by Druckman and Jackson (2010). The decent living scenario has emissions that are just under the 4.1–6.7 tonnes per capita range predicted by Lamb and Rao (2015), while the good life scenario has emissions that are above it. Huo et al. (2023) arrive at a much lower figure of 2.3 tonnes per capita, in part by basing their model on the supply chains of emerging economies instead of a developed country such as the UK.

The energy footprint of our decent living scenario (35 GJ per capita) is consistent with, but on the high side, of previous studies that estimate minimum energy footprints, which range from 9 to 45 GJ per capita per year (Millward-Hopkins et al., 2020; Rao et al., 2019; Grubler et al., 2018; Lamb and Rao, 2025; Kikstra et al., 2021). The energy footprint of the good life scenario (73 GJ per capita) is substantially higher than the range of these previous studies, however. The material footprint of the decent living scenario (5.5 tonnes per person) is within the 2–11 tonnes per person range given by Vélez-Henao and Pauliuk (2023), and very near their reference value of 5 tonnes per person. The good life scenario is higher than this range, at 13.2 tonnes per person.

Overall, our results for energy and emissions are a bit higher than previous studies, and this difference likely stems from two main factors — our use of the UK's technology and supply chains as the basis of our scenarios, and the top–down approach of MRIO analysis. With respect to technology, our analysis does not consider cleaner energy production, or lifestyle changes such as a shift to plant-based diets, all of which could lower the footprints. The minimum income standard that we build upon in our analysis is explicitly designed to reflect the reality of UK society in 2019. For example, families with children were allocated cars because public transit was perceived as not being accessible in the focus groups (Davis et al., 2012). With respect to the methodological approach, Millward-Hopkins et al. (2020), Rao et al. (2019), and Grubler et al. (2018) all use a bottom–up methodology. While a bottom–up approach allows for modelling more radical changes to provisioning systems, it can fail to include many of the indirect or implicit resource requirements that MRIO accounts for, such as infrastructure, capital investment, and governance. Notably, the gaps in Rao and Min's (2018) decent living standards outlined in Section 5.1 suggest that these previous estimates are likely missing many essential components that are captured in the good life scenario. Our analysis, however, is unable to model the radical changes to the UK economy that Rao and Min (2018) imagine.

Despite methodological differences, our results for the decent living scenario are relatively close to previous results measuring the energy, emissions, and material footprints of the decent living standard. The requirements of the good life scenario, in contrast, are about twice those of the decent living standard and higher than previous results in the literature. This finding suggests that the factors separating the good life scenario from the decent living scenario — such as government functions, the education of skilled workers, care work, household goods, maintenance of built capital, and recreation — require substantial labour, energy, emissions, and materials, which should not be overlooked.

## 5.4 Implications

Some of our findings may seem surprising. We expected that limiting consumption to basic needs would result in a low and socially sustainable labour footprint, similar to Keynes' prediction of a 15-hour work week. While this is true for the decent living scenario, the good life scenario requires increased labour, energy, emissions, and material use compared to the global average, and is environmentally unsustainable.

While the decent living scenario has a low labour footprint and approaches environmental sustainability, it underestimates what is required for human wellbeing by excluding important societal needs. The good life scenario addresses these gaps based on what UK residents say they require for a good life, but this approach may lead to overestimates in some categories. For instance, the categories for care work, recreation, and gross fixed capital formation add 18.5 hours per week to the labour footprint of the good life scenario compared to the decent living scenario, in effect doubling the latter.

Childcare and recreation are clearly important, but their inclusion as economic sectors in the good life scenario could also be seen as a form of commodification. The time required for childcare is not counted in the decent living scenario, but this does not mean it does not exist, i.e. as unpaid work.

Future research could try to account for unpaid work, which might bring the two scenarios closer together.

The good life scenario also includes the spending required to maintain the stock of built capital, which is clearly important for human wellbeing. However, there are undoubtedly components of the capital stock that are not needed for a good life (e.g. fossil fuel infrastructure), whose maintenance costs could be subtracted, while there are other components (e.g. renewable energy) where expenditure should be expanded. Our study draws attention to the importance of the capital stock in meeting basic needs, but future research could refine this calculation further.

Overall, our results suggest that limiting consumption to the level of basic needs is not enough on its own to achieve sustainability. Instead, we argue that achieving sustainability requires a two-pronged approach. The first prong is limiting consumption to the level of meeting basic needs (sufficiency), while the second prong is improving provisioning systems (efficiency).

Regarding sufficiency, countries like the UK, where labour and resource use exceed basic needs, could reduce consumption to this level. Instead of pursuing GDP growth, wealthy nations could pursue a "post-growth" economy, where the aim is to improve human wellbeing while reducing resource use (Kallis et al., 2025; Slameršak et al., 2024). The "Doughnut" of social and planetary boundaries provides a useful framework for measuring progress towards such an economy, by quantifying social outcomes in comparison to thresholds for basic needs, and resource use in comparison to planetary boundaries (Raworth, 2017; O'Neill et al., 2018). Policies proposed to help achieve a post-growth economy include universal basic services, working-time reduction, a maximum income, carbon taxes, and measures of progress beyond GDP (Kallis et al., 2025; Vogel et al., 2024).

Efficiency, in a broad sense, could be improved through both social policies (such as reducing inequality) and technologies (such as public transit). Provisioning systems — which mediate the relationship between resource use and social outcomes (Fanning et al., 2020) — encompass both social structures (including government institutions, cultural norms, and markets) and physical systems (such as infrastructure and technology).

Our static MRIO analysis does not explore improvements that could be made to provisioning systems, but glimpses are still apparent in the data and literature. For instance, India has less than an eighth of the per capita emissions of the US, but it does not have one eighth of its life expectancy or happiness. Rao and Min (2018) found that healthcare funding above PPP $700 per capita had little impact on life expectancy, while Fanning and O'Neill (2019) found no significant relationship between changes in GDP and life expectancy. While the UK's national health service could not function in its current form with only 13% of its funding (as in the decent living scenario), there may be ways to provide similar levels of care with a lower budget, such as a greater focus on preventive medicine or healthy lifestyles (Campbell, 2015; Heron et al., 2019). Vogel et al. (2021) have found that higher need

satisfaction can be achieved at lower energy use through more public services, greater income equality, and stronger democracy.

Finally, the role of technological change is also important to consider. Advances in artificial intelligence may increase labour productivity (Aghion and Bunel, 2024), and therefore reduce the amount of work required to meet basic needs. However, historical improvements in labour productivity and resource efficiency have often not translated into reductions in working hours or resource use (Hayden and Shandra, 2009). One reason for this is the rebound effect or "Jevons paradox" (Polimeni et al., 2008). Briefly stated, improvements in efficiency tend to lead to a reduction in costs, which in turn leads to an increase in production and consumption. Strict caps on resource use, emissions, and working time may be needed to break this cycle, directing growth in productivity to increasing leisure time, rather than to increasing production and consumption.

## 5.5 Novel contributions

This article makes several important contributions to the literature. It presents the first comprehensive calculation of the labour and resources required to provide a good life for all, based on the minimums demanded by the population of a developed country. The specific results are both novel and challenge previous findings in the literature.

Beyond its results, the article also makes important methodological contributions. It contributes a new set of basic needs categories (Table 1), which go beyond the household-focussed decent living standards of Rao and Min (2018) to include additional forms of consumption that are required to maintain society (e.g. government, recreation, care work, and maintenance of the capital stock). In practical terms, it provides concordance tables linking these categories to the minimum income budgets of Hirsch (2019), the decent living standards of Rao and Min (2018), and the 200 sectors in EXIOBASE 3 (see Supplementary Materials 2).

Lastly, it provides a new unit of account for labour footprints (hours per week equivalent), which maintains the scalability and tangibility of per capita accounts, while accounting for the number of years people are likely to work during their life and the number of working weeks per year. This unit allows for easy comparison between results for different populations or different years, as well as a more intuitive understanding of what the results mean in practice.

## 5.6 Limitations and future research

The approach we have adopted in this study also has limitations. While MRIO analysis is a powerful method for estimating the paid labour and resource use associated with different consumption profiles, it functions at the sectoral level. The method is particularly well suited to spending categories that align directly with economic sectors, such as education and healthcare. However, it is less well suited to spending categories linked to specific products, such as appliances or mobility. Graeber (2018) argues that modern economies contain a substantial number of “bullshit jobs”. These are jobs that are not needed to meet people’s needs, but which serve the role of keeping people plugged into the economic system. Given that our analysis explores the impact of reducing consumption on a sector-by-sector basis, it is not able to capture this effect. Moreover, MRIO is a static methodology, and does not account for changes over time. Potential changes in prices, living standards, technology, or labour productivity that could occur in the future are therefore not accounted for in our analysis. Nor is the important contribution of unpaid labour.

The method is also limited by the underlying data. We rely heavily on the spending profiles provided by Hirsch (2019). These spending profiles are based on UK residents’ opinions about minimum standards. Rao and Min (2018) suggest that the UK minimum income standard reflects cultural embeddedness more than universal needs. As discussed in Section 5.2, both the social standard of the UK and its existing supply chains are reflected in our results. The limitations of the minimum income standards are recognised by the Joseph Rowntree Foundation (Bradshaw et al., 2008; Davis et al., 2012; Davis et al., 2018). These include unrealistically high budgets in some cases, and a lack of accounting for people with disabilities, particular health or diet needs, or little access to urban resources.

Future research could help overcome these limitations. More comprehensive bottom–up analyses of labour requirements could help establish a lower bound for the labour required to provide a good life for all. Time use surveys could be used to estimate the contribution of unpaid labour. Further MRIO analysis could use other countries’ supply chains and minimum social standards, drawing on the work of Ng et al. (2023), Aban Tamayo et al. (2021), or other studies referenced by Loughborough University (2024). Such an analysis could also attempt to account for future changes in economic structure, more localised supply chains, technological improvements (e.g. artificial intelligence), and demographic changes.

# 6. Conclusion

This article has explored the labour, energy, emissions, and material footprints of providing a good life for all people. A bare bones decent living scenario, based on the decent living standards of Rao and Min (2018), would require a 17.5-hour working week, and on a per capita basis, 35 GJ of energy use, 4.0 tonnes of emissions, and 5.5 tonnes of material use. While this scenario approaches environmental sustainability, it fails to meet important societal needs. The more socially sustainable good life scenario, based on minimum social standards demanded by UK residents, would require an unsustainably high 45.5-hour working week, 73 GJ of energy use, 7.5 tonnes of emissions, and 13.2 tonnes of material use per capita.

Our results also reveal the imperialist appropriation embodied in the social customs, laws, and economy of the UK. They illustrate how the opulence of developed countries requires continued extraction of labour and resources from some of the poorest regions in the world. The decent living scenario requires unconscionable cuts to UK society, while the good life scenario requires surprisingly high working hours. Both scenarios involve significantly more greenhouse gas emissions than the 2 °C carbon budget allows for. Neither of these low-consumption scenarios represent a realistic path to providing a good life for all.

Overall, our results suggest that limiting consumption to the level of basic needs is likely a necessary, but not sufficient, condition for social and environmental sustainability. If limiting consumption is not enough on its own to achieve a good life for all people, then radical changes to provisioning systems are also required. Our analysis suggests that the provisioning systems and supply chains of the UK are less labour and resource efficient than those of many other countries, and could likely be improved, although cultural norms would have to shift as well. Ultimately, providing a good life for all people requires following a very different path than we are on today.

# Appendix
# Detailed Scenario Descriptions

Our analysis focuses on two low-consumption scenarios: a decent living scenario and a good life scenario. Each scenario is defined through 13 spending categories. In each category, several data sources are integrated to generate a total budget, which we then used as an input for multi-regional input–output analysis. This document includes detailed descriptions of the methods used to calculate budgets for all spending categories in both scenarios.

## A1. The decent living scenario

The decent living scenario matches the household and collective requirements of the decent living standard proposed by Rao and Min (2018). Rao and Min (2018) include 12 household requirements and 10 collective requirements. These material requirements were converted into spending budgets largely based on the minimum budgets provided by Hirsch (2019), which are explained in detail by Bradshaw et al. (2008) and several other sources. While these minimum budgets were used as a starting point, they were reduced in almost every category to match the decent living standards proposed by Rao and Min (2018). The process to generate the personal budget for each category is explained in detail below.

### A1.1 Groceries (food and drinks)

The first household requirement is a minimum daily intake of calories, protein, vitamins, and minerals, which was paired with the food and beverage categories of EXIOBASE and the food category of the Hirsch (2019) budgets. Since Rao and Min (2018) explicitly exclude alcohol from their requirements, it is not included in the decent living scenario. Hirsch (2019) includes the cost of dining out in the food budget despite discussing it in the social participation budget. Based on the detailed budgets provided by Loughborough University (2023b), dining out represented between 10.1% and 13.8% of the provided food budget in 2018, averaging around 11.6%. To account for this, 11.6% of the Hirsch (2019) food budget was therefore removed for the decent living scenario.

### A1.2 Clothing

Rao and Min's (2018) clothing requirement is aimed at achieving basic comfort, while the Hirsch (2019) budgets seem more focused on social acceptability (Bradshaw et al., 2008). However the standards were deemed similar enough, and the Hirsch (2019) clothing budget was left untouched.

### A1.3 Housing

For housing, Rao and Min (2018) recommend 30 $m^2$ per household, with only 10 $m^2$ more area for each person after the third. As every housing type except for multi-family housing averages less than four people per household, the 37 $m^2$ area of an average 1-bedroom, 1-bath flat (Roberts-Hughes, 2011) was deemed a good average for all households in the decent living scenario. The Hirsch (2019) budget includes two types of 1-bedroom flats: social housing and privately rented. It was deemed simpler to use the privately rented cost for all households, and exclude the cost of housing from public spending.

A1.4 Utilities and insurance

To ensure acceptable living conditions, including a safe space, basic comfort, and hygiene, Rao and Min (2018) include heating and cooling equipment, electricity, toilets, waste management, and water supply as requirements. These requirements are well-matched by the water rates and fuel budget from Hirsch (2019), which includes electricity and gas for residences, but not fuel for transit. Insurance is not discussed by Rao and Min (2018) and therefore is not included.

A1.5 Healthcare

Healthcare spending was reduced to meet the PPP $700 per capita threshold specified by Rao and Min (2018). This threshold represents a large (87%) decrease from real healthcare spending in the UK in 2019, but represents the upper end of Rao and Min's (2018) PPP $450–700 per capita range.

A1.6 Appliances, furnishing, and maintenance

Rao and Min (2018) include provisions for a refrigerator, heating/cooling equipment, shared washing machines, adequate lighting, toilets, and cook stoves. For privately rented flats, which all households have in this scenario, Hirsch (2019) assumes that these items are provided as part of rent, with the sole exception of light bulbs, which Loughborough University (2023b) prices at about £0.06 per week per household in 2018. Light bulbs were deemed a trivial expense and excluded from the calculations, so this category was left at £0 for the decent living scenario. This means that items like mattresses, eating/cooking utensils, chairs, lampshades, and cleaning equipment, which are not mentioned by Rao and Min (2018), were not included in the scenario.

A1.7 Education

Rao and Min (2018) only require 9 years of education in their decent living standards. This value would include primary education and roughly half of secondary education in the UK. All public spending for primary education and half of the spending on secondary education represents 54.6% of UK public spending on education (HM Treasury, 2019). Since private funding data for education in the UK are largely unavailable (Elaine et al., 2023), public funding was deemed an acceptable proxy for overall spending, and spending on education was reduced to 54.6% of UK public spending on education for the decent living scenario.

A1.8 Devices (TVs, phones, computers)

Rao and Min (2018) include one TV for each household, one phone either per household or per adult (there are inconsistencies), and communication infrastructure. Hirsch (2019) includes a TV, TV license, laptop, landline and mobile phones, cell and internet service, and in some instances a radio and DVD player. The Hirsch (2019) budget was therefore reduced to only include a TV, TV license, mobile phone, and phone service, based on the 2018 numbers provided by Loughborough University (2023b). No personal computers were included.

A1.9 Transport

The allotments for transport are quite similar between Rao and Min (2018), who require public transport in urban areas and some private vehicles for rural areas, and Hirsch (2019), who includes funding for public transport, bikes, and occasional taxis for most households, and private cars for families with multiple children. As such, the Hirsch (2019) budgets were left untouched.

A1.10 Public administration and defence

Public administration and defence spending is included as a single sector in EXIOBASE, and was scaled by overall spending as described in Section 3.3 of the main manuscript. The scaling was based on the ratio between total public spending in the UK in 2019 and the hypothetical total had spending been limited to the categories described in Rao and Min's (2018) decent living standards (HM Treasury 2019, pp. 75–77). We did not include spending for general public services, defence, public order and safety, and environmental protection, as none of these are included as collective requirements by Rao and Min (2018). Economic affairs spending was limited to the proportion used for fuel and energy; mining, manufacturing, and construction; transport infrastructure; and ICT infrastructure. Housing and community amenities spending was restricted to only include community development, water supply, and street lighting. Education, healthcare, recreation, and waste management are covered by other spending categories and therefore not included in this category.

A1.11 Recreation (vacations, toys, dining out)

Spending on recreation was set to zero, as Rao and Min (2018) make no mention of vacations, toys, or dining out. They mention leisure several times, including traveling for leisure, but no specific allowances for these activities are made, such as hotels, long distance travel, or leisure goods.

A1.12 Care work (childcare, senior care)

Spending on care work was also set to zero, as Rao and Min's (2018) discussion only includes it as an aspect of healthcare. Senior care and childcare are explicitly mentioned as aspects of healthcare, and are assumed to be covered under the total PPP $700 per capita healthcare budget.

### A1.13 Gross fixed capital formation

Finally, spending on gross fixed capital formation (GFCF) was also set to zero. Rao and Min (2018) make no allowances for depreciation or the investment necessary to maintain industrial equipment. For consistency with other results based on their decent living standards, spending on GFCF was not included in this scenario.

## A2. The good life scenario

This scenario is based largely on the minimum income standards published by the Joseph Rowntree Foundation (Bradshaw et al., 2008; Hirsch, 2019). The budgets determined by Hirsch (2019) were used for every category except for care work and healthcare, and the three added categories (education, public administration, and gross fixed capital formation). Per-category spending is explained in detail below.

### A2.1 Groceries (food and drinks)

This category includes food and drinks (including alcohol) bought from grocery stores. Food was selected for nutritional adequacy and affordability (Bradshaw et al., 2008). Restaurants and bars are not included. However, as mentioned in Section 1.1, Hirsch (2019) includes the cost of dining out in the food budget despite discussing it in the social participation budget. Therefore the same 11.6% reduction was applied here, and added to the recreation budget.

### A2.2 Clothing

Clothing was selected for a range of wearers and the budget was based on product lifetimes estimated by groups of UK residents (Bradshaw et al., 2008).

### A2.3 Housing

Expected costs in the Hirsch (2019) budgets are based on average social rents in the East Midlands, for all households except working age adults without children, who were expected to live in privately rented apartments (Davis et al., 2018). Social rent prices are lower than private rent due to public subsidies. The good life scenario assumes that everyone would be living in apartments with rents similar to social housing, including working age adults without children, to allow for more straightforward calculations. To estimate the total cost, spending on social housing was increased by a factor of 5.9, proportional to the increase in overall social housing that this would represent, from 17% to 100% of households (Ministry of Housing, Communities and Local Government, 2021). However, this spending was allocated to the housing category instead of the public administration category.

A2.4 Utilities and insurance

This category includes water rates, household insurances, and costs for residential power and natural gas use (the last two of which Hirsch, 2019 refers to as "fuel").

A2.5 Healthcare

Hirsch (2019) includes a small allocation for personal healthcare needs, under the title "personal goods and services". This allocation was replaced with a more comprehensive account of healthcare based on total domestic spending on healthcare in the UK in 2019 by households, the government, and non-profits (provided in the EXIOBASE data). A 30% reduction from this total spending amount was introduced based on the repeated finding that about 20–40% of healthcare spending is wasted (WHO, 2014; Berwick and Hackbarth, 2012).

A2.6 Appliances, furnishing, and maintenance

This category includes two categories from Hirsch (2019): "other housing costs" and "household goods". The former includes a budget for decorations and maintenance, and the latter includes a large variety of items, such as "lampshades and light bulbs, carpets and curtains, kitchen appliances, and all furniture, cooking utensils, crockery, electrical devices, textiles, etc" (Bradshaw et al., 2008, p. 20). As working age households without children are not living in private flats in this scenario, the price of a fridge, stove, and washing machine were added back in based on the costs given by Loughborough University (2023b).

A2.7 Education

As a considerable amount of both new research and skill development seems necessary to support this scenario, all existing education (elementary, secondary, and tertiary) was included in the good life scenario. To model this, the education budget for the good life scenario was set equal to the baseline education spending in the UK in 2019.

A2.8 Devices (TVs, phones, computers)

This category was assembled from pieces of two categories from the Hirsch (2019) budgets. The first category was the "household services" category, which includes funding for a landline and mobile phone, as well as home internet. It also includes a fairly small cost for stamps for paper mail (Loughborough University, 2023b). This category was therefore included in full. The remaining devices were pulled from the "social participation" category, using 2018 costs from Loughborough University (2023b). The specific calculations can be seen in the "Hirsch (2019) spending" spreadsheet in Supplementary Materials 2.

A2.9 Transport

Transport is combined from two Hirsch (2019) categories: “motoring”, which includes the costs of owning and operating a car, and “other travel costs”, which covers taxi, bus, and rail fares (not costs of long-distance transit for leisure) (Davis et al., 2018). Public transport was deemed too expensive and inconvenient for families with children, who were given the budget for a private car to get by in urban areas (Davis et al., 2018). Families without children had no motoring costs.

A2.10 Public administration and defense

The good life scenario includes over three times the public administration budget of the decent living scenario. This budget does not include interest payments on public debt, spending on social protection, or EU transactions, as these don’t show up as final government spending in EXIOBASE. Education, healthcare, waste management, and recreation are also not included, because they are represented in other, more specific sectors in EXIOBASE. This leaves 9 broad categories: executive and legislative functions, defense, public order and safety, economic affairs (which includes industry regulations and participation in particular industries), transportation, utilities, environmental protection, public spaces, and housing. All of these categories were included in the good life budget except for defense, which seemed counter to an economy focused on achieving a good life for all people globally. Defense was about 28% of baseline spending in this category.

A2.11 Recreation (vacations, toys, dining out)

This category is the remainder of social participation, once spending on electronic devices is removed. It includes toys and pocket money for children, group sports, dining out (about four times a year for families and every two weeks at cheaper places for pensioners), and holiday travel (Davis et al., 2018). The 11.6% of the food budget that represents dining out was moved here, as explained in Section 2.1. Holiday travel includes a one-week per year holiday within the UK (Davis et al., 2018).

A2.12 Care work (childcare, senior care)

The care work budget was set to 20% of the Hirsch (2019) budget because this budget assumes that all of a child’s parents work full time (Bradshaw et al., 2008). In practice only 16% of children in couple-parent families, and 22% of children in lone-parent families, are in this situation (Bradshaw et al., 2008). Senior care is assumed to be covered under healthcare.

A2.13 Gross fixed capital formation

Gross fixed capital formation (GFCF) is included as a separate spending vector in EXIOBASE. It includes both capital investment for increases in production and spending to cover depreciation. The latter category is considered a necessary cost of production, and included in the good life scenario. The amount of GFCF spending necessary to cover depreciation of assets is estimated as 14.9% of GDP, based on US figures for investment and GDP (FRED, 2024a; 2024b; 2024c).